\documentclass[a4paper]{article}

\usepackage[dvips]{graphicx}
\usepackage{algorithm}
\usepackage{algpseudocode}
\usepackage{cite}
\usepackage{amssymb}
\usepackage{hyperref}

\usepackage[latin1]{inputenc}
\usepackage{tikz}
\usetikzlibrary{shapes,arrows,positioning,calc}

\tikzstyle{startstop} = [rectangle, rounded corners, minimum width=3cm, minimum height=1cm,text centered, draw=black, fill=red!30]
\tikzstyle{io} = [trapezium, trapezium left angle=70, trapezium right angle=110, minimum width=3cm, minimum height=1cm, text centered, draw=black, fill=blue!30]
\tikzstyle{process} = [rectangle, minimum width=3cm, minimum height=1cm, align=left, draw=black, fill=orange!30]
\tikzstyle{empty} = [rectangle, align=left]
\tikzstyle{decision} = [diamond, minimum width=3cm, minimum height=1cm, align=center, draw=black, fill=green!30]
\tikzstyle{arrow} = [thick,->,>=stealth]

\begin{document}
%
\title{\flushleft \bf Modeling of screening currents in coated conductor magnets containing up to 40000 turns\footnote{\color{blue}The final version of the article can be found in the publication only: Supercond. Sci. Technol., vol. 29, a.n. 085004 (2016), \url{http://dx.doi.org/10.1088/0953-2048/29/8/085004}.}}
\markboth{Modeling of screening currents in coated conductor magnets ...}{}

\author{E. Pardo}%

\author{Enric Pardo\\
{\normalsize Institute of Electrical Engineering, Slovak Academy of Sciences,}\\
{\normalsize Dubravska 9, 84104 Bratislava, Slovakia}\\
{\normalsize enric.pardo@savba.sk}
}

\date{\today}

\maketitle

\begin{abstract}

Screening currents caused by varying magnetic fields degrade the homogeneity and stability of the magnetic fields created by $RE$BCO coated conductor coils. They are responsible for the AC loss, which is also important for other power applications containing windings. Since real magnets contain coils exceeding 10000 turns, accurate modeling tools for this number of turns are necessary for magnet design. This article presents a fast numerical method to model coils with no loss of accuracy. We model a 10400-turn coil for its real geometry and coils of up to 40000 turns with the continuous approximation, which we checked that introduces negligible errors. The screening currents, the Screening Current Induced Field (SCIF) and the AC loss is analyzed in detail. The SCIF is maximum at the remnant state with a considerably large value and decreases substantially after one day of relaxation. The instantaneous AC loss for an anisotropic magnetic-field dependent $J_c$ is qualitatively different than for a constant $J_c$, although the loss per cycle is similar. Saturation of the magnetization currents at the end pancakes causes that the maximum AC loss at the first ramp increases with $J_c$. The presented modeling tool can accurately calculate the SCIF and AC loss in practical computing times for coils with any number of turns used in real windings, enabling parameter optimization.

\end{abstract}



\section{Introduction}
\label{s.intro}

Time varying magnetic fields create screening currents in superconductors \cite{reviewAC}. The screening currents induced at the charge and discharge ramps of magnets distort the generated magnetic field \cite{maeda14IES}, being this critical for magnet applications like Magnetic Resonance Imaging (MRI), Nuclear Magnetic Resonance (NMR) and accelerator magnets. In addition, screening currents create dissipation in the winding, which increases the running costs and complicates the cryogenics. This is also relevant for other applications containing windings, such as Superconducting Magnetic Energy Storage (SMES), transformers and rotating machines. Screening currents in $RE$BCO\footnote{$RE$BCO stands for $RE$Ba$_2$Cu$_4$O$_{7-x}$, where $RE$ is a rare earth, usually Y, Gd or Sm.} coated conductors, with relatively high width, are of higher importance than in multi-filamentary wires, such as NbTi, Nb$_3$Sn or MgB$_2$. In addition, superconducting magnets (and SMES) may contain thousands of turns \cite{weijers14IES,wangQ15IES}, reaching up to around 30000 turns \cite{awaji14IES}. Therefore, modelling of the screening currents in superconducting windings of high number of turns is necessary for the design of these applications.

The interaction between screening currents strongly impact their magnitude, as has been found for stacks of tapes in external applied magnetic field \cite{mawatari96PRB,pardo03PRB,grilli06PhC} and coils \cite{neighbour,luJ15IES}. Modelling of the screening currents taking their interaction between all turns into account have been published for pancake coils (or pancakes) with more than 100 turns \cite{pancaketheo,prigozhin11SST,zermeno11IES,ainslie12PhC} and stacks of pancakes with up to 768 turns \cite{pardo12SSTb,zhangM14SST}. In order to model higher number of turns with practically no loss in accuracy, the continuous approximation has been developed \cite{prigozhin11SST,neighbour,zermeno13JAP}. This allowed to model coils of up to 4000 turns \cite{pardo15SST,xiaJ15SST,queval16SST,amemiya16SST}. Predecesor models of the continuous approximation have been published in \cite{Clem07SSTb,yuanW10JAP}, which take additional approximations. Semi-analytical models have also been applied to model magnet size coils with up to around 30000 turns, under certain approximations \cite{kajikawa14arxiv}. Coils of around 900 turns have been modeled by a 3D numerical method, which intrinsically assumes the thin film geometry of the tapes \cite{ueda15IES}.

This work presents a method to accurately calculate the screening currents in coils containing more than 10000 turns (one order of magnitude higher than previous works) with feasible computing times\footnote{The essence of the numerical model in this article and part of the results have been presented in conferences between 2014 and 2015 \cite{pardo14HTSmod,pardo15MT}; presentations are available on-line in \cite{pardo14HTSmod,kapolka15EUCAS}.}. By means of the continuous approximation, the number of turns is extended up to 40000. This fulfills the present requirements for high-field magnets and other applications containing windings. When using the real geometry, the numerical method does not assume that the tape is a thin film; and hence current penetration across the thickness can be modeled, which is important for coils made of powder-in-tube tapes like Bi2223.

By means of this method, we calculate the Screening Current Induced Field (SCIF) at the center of a 10400-turn coil for both a cyclic ramp-up and ramp-down process and relaxation after the end of the first ramp down, or remnant state. The article also presents the SCIF at the remnant state for coils from 800 to 40000 turns. We analyze in detail the instantaneous AC loss in the 10400 turn coil for cyclic ramp-up and ramp-down. Although this work regards the coil as a stand-alone magnet, adding the effect of the magnetic field generated by an outsert is straightforward.

\begin{figure}[tbp]
\centering
{\includegraphics[trim=40 30 25 46,clip,width=8.5 cm]{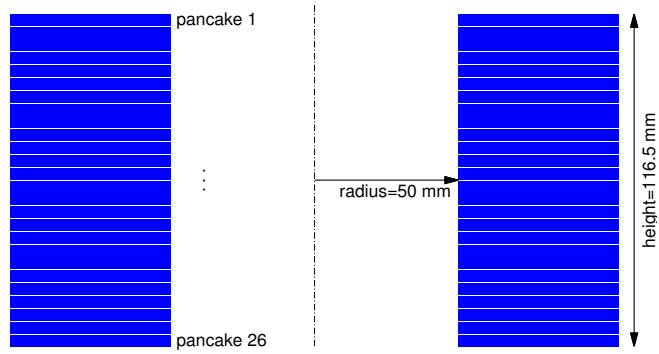}}
\caption{ Sketch of the cross-section of the coil of the main analysis, containing 10400 turns. Computations for other coils only differ in the total number of pancakes. }
\label{f.sketch}
\end{figure}

\begin{figure}[tbp]
\centering
{\includegraphics[trim=6 84 8 75,clip,width=8.5 cm]{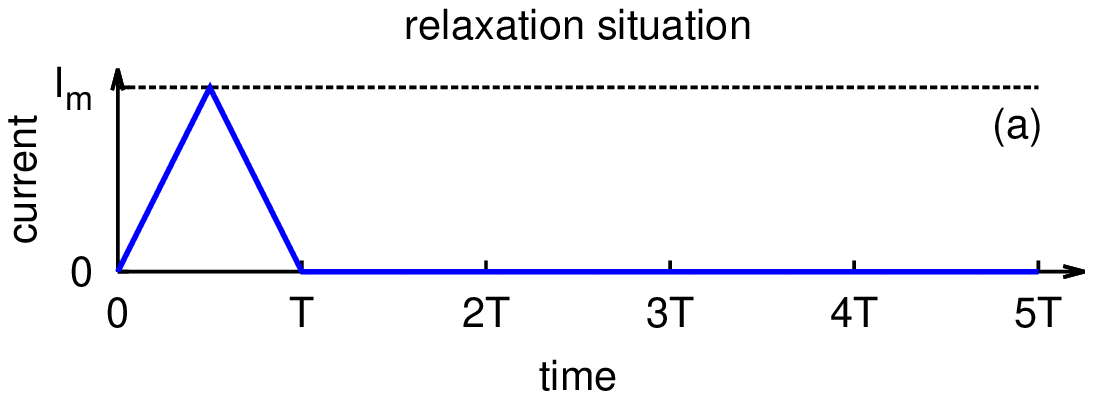}}
{\includegraphics[trim=6 53 8 75,clip,width=8.5 cm]{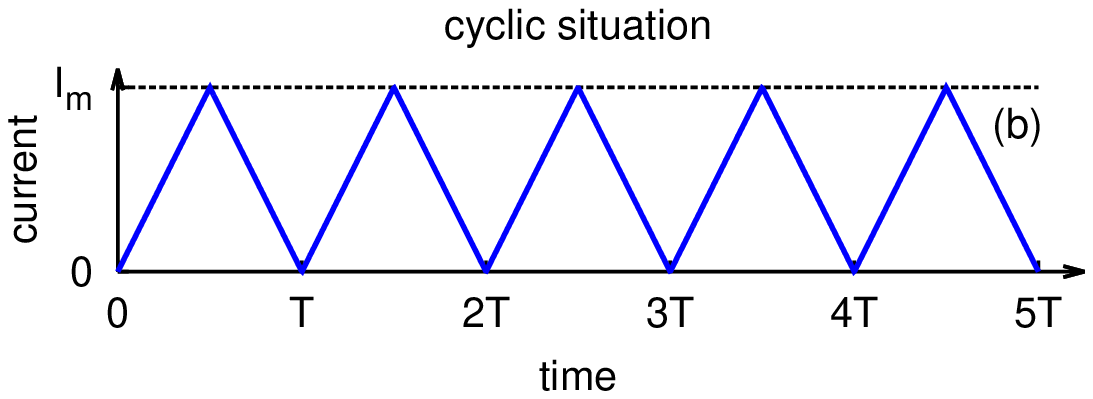}}
\caption{ Qualitative time signal of the current for (a) remnant state relaxation and (b) cyclic excitation. }
\label{f.signal}
\end{figure}


\section{Studied geometry}

\begin{table}
\begin{center}
\begin{tabular}{ll}
\hline
\hline
number of pancakes & 26 \\
number of turns per pancake & 400 \\
total number of turns & 10400 \\
inner radius & 50 mm \\
outer radius & 105.9 mm \\
superconducting layer thickness & 2 $\mu$m \\
superconducting layer width & 4 mm \\
radial separation between superconducting layers & 138 $\mu$m \\
axial separation between pancakes & 0.5 mm \\
\hline
\hline
\end{tabular}
\caption{ Parameters of the coil of the main analysis. Calculations for other coils only differ in the total number of pancakes. }
\label{t.geom}
\end{center}
\end{table}

This article analyzes in detail the winding with parameters in table \ref{t.geom}, which contains 10400 turns and its cross-section is in figure \ref{f.sketch}. In section \ref{s.remIc}, We also present the Screening Current Induced Field (SCIF) at the remnant state and the critical current for coils with the same parameters as in table \ref{t.geom} but 2 to 100 pancakes (800 to 40000 turns).


\section{Model}
\label{s.method}

\begin{figure}[tbp]
\centering
{\includegraphics[trim=0 0 0 0,clip,width=8.5 cm]{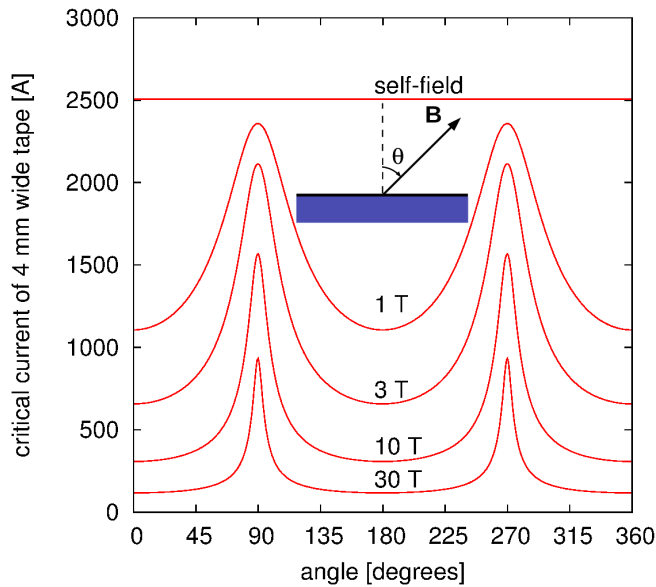}}
\caption{Anisotropic magnetic field dependence of the tape critical current, $I_{ct}(B,\theta)$, as input for the calculations, given by equations (\ref{IcBth}) and (\ref{wB}) \cite{hilton15SST}. We assume that the local $J_c$ follows $J_c=I_{ct}/(wd)$, where $w$ and $d$ are the width and thickness of the superconducting layer, respectively.}
\label{f.Ict}
\end{figure}

We calculate the screening currents by means of the Minimum Magnetic Entropy Production (MEMEP) method \cite{pardo15SST}. This numerical method computes the detailed current density in each turn for any given $E(J)$ relation, where $E$ and $J$ are the electric field and current density, respectively. In axially symmetric geometries, $J$ is the angular component of the current density. In this article, we assume
\begin{equation}
E(J)=E_c\left( { \frac{|J|}{J_c} } \right)^n\frac{J}{|J|},
\end{equation}
where $J_c$ is the critical current density, $n$ is the power-law exponent and $E_c=10^{-4}$ V/m. We took the dependence of $J_c$ on the magnetic field\footnote{In this article we refer to the magnetic field (or ``field"), ${\bf H}$, and magnetic flux density, ${\bf B}$, indistinctly because magnetic materials are not present, and hence ${\bf B}=\mu_0{\bf H}$.} $B$ and its angle with the normal of the tape surface $\theta$ from \cite{hilton15SST}, based on fitting experiments of a 4 mm wide SuperPower tape \cite{SuperPower} at 4.2 K from self-field to 30 T. Using the angle $\theta$ definition of figure \ref{f.Ict}, the fit for the tape critical current in \cite{hilton15SST} is 
\begin{eqnarray}
I_{ct}(B,\theta) & = & \frac{K_0}{(\frac{B}{\beta_0}+1)^{a_0}}+\frac{K_1}{(\frac{B}{\beta_1}+1)^{a_1}}\cdot \nonumber \\
&& \cdot[\omega^2(B)\cos^2(\theta-\theta_0)+\sin^2(\theta-\theta_0)]^{-1/2} \label{IcBth}
\end{eqnarray}
with
\begin{equation}
\omega(B)=\left ( { \frac{B}{\beta_\omega}+1 } \right )^{5/3} \label{wB}
\end{equation}
where $B$ is the applied magnetic field and we re-wrote the equations above in order to get constants with integer dimensions, with values $K_0=292.5$ A, $K_1=2213$ A, $\beta_0=13.8$ T, $\beta_1=13.8$ T, $\beta_\omega=0.2792$ T, $a_0=1.3$, $a_1=0.809$ and $\theta_0=-0.180^{\rm o}$. At high magnetic fields, this $I_{ct}(B,\theta)$ dependence presents a sharp peak at $\theta=90.18^{\rm o}$ and a wide plateau around $\theta=0^{\rm o}$ (figure \ref{f.Ict}). In this article, we assume that $J_c(B,\theta)=I_{ct}(B,\theta)/(wd)$, where we take $B$ as the local magnetic field and $w$ is the tape width and $d$ is the thickness of the superconducting layer. Then, we assume that the tape self-field is negligible compared to the total magnetic field in the magnet, at least in the regions where $J$ is limited by $J_c$.

\begin{figure}[tbp]
\centering
{\includegraphics[trim=0 0 0 0,clip,width=8.5 cm]{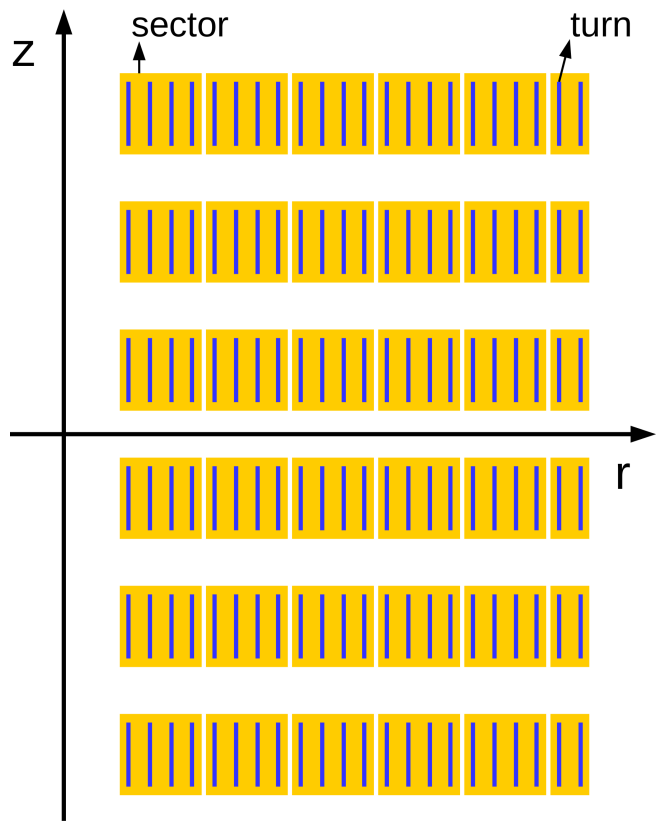}}
\caption{Dividing the coils into sectors enables to speed up the computations and efficiently parallelize the numerical method. }
\label{f.sectors}
\end{figure}

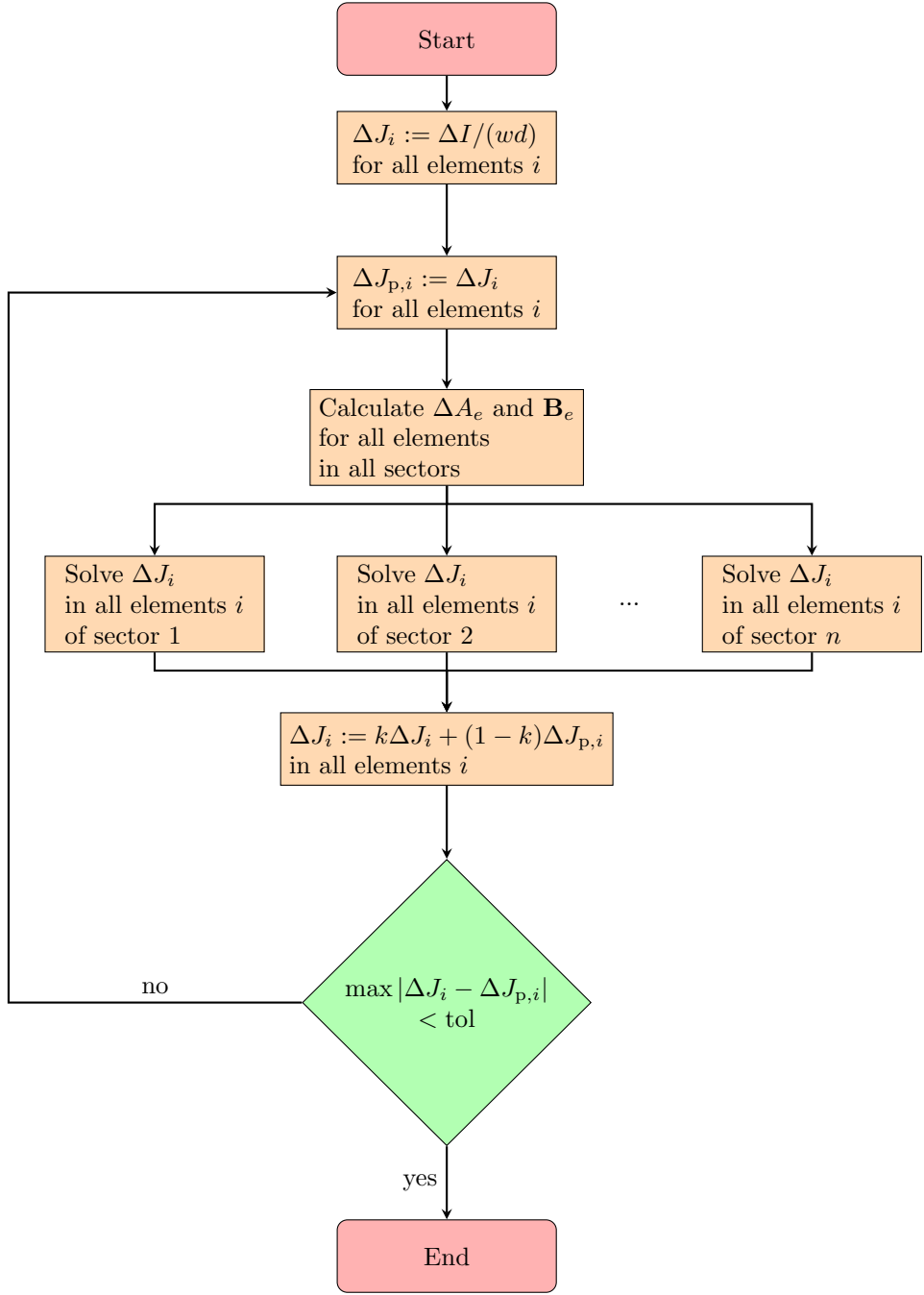
\begin{figure}
\begin{tikzpicture}[node distance = 2cm, auto]
	\node (start) [startstop] {Start};
	\node (Jinit) [process, below of=start, node distance=1.5 cm] {$\Delta J_i:=\Delta I/(wd)$\\for all elements $i$};
	\node (Jp) [process, below of=Jinit] {$\Delta J_{{\rm p},i}:=\Delta J_i$\\for all elements $i$};
	\node (A0B0) [process, below of=Jp] {Calculate $\Delta A_e$ and ${\bf B}_e$\\for all elements\\in all sectors};
	\node (sec2) [process, below of=A0B0,node distance=2.3 cm] {Solve $\Delta J_i$\\in all elements $i$\\of sector 2};
	\node (sec1) [process, left of=sec2, node distance=4 cm] {Solve $\Delta J_i$\\in all elements $i$\\of sector 1};
	\node (dots) [empty, right of=sec2, node distance=2.5 cm] {...};
	\node (secn) [process, right of=dots, node distance=2.5 cm] {Solve $\Delta J_i$\\in all elements $i$\\of sector $n$};
	\node (damp) [process, below of=sec2] {$\Delta J_i:=k\Delta J_i+(1-k)\Delta J_{{\rm p},i}$\\in all elements $i$};
	\node (tol) [decision, below of=damp, node distance=3.5 cm] {$\max|\Delta J_i-\Delta J_{{\rm p},i}|$\\$<{\rm tol}$};
	\node (end) [startstop, below of=tol, node distance=3.5 cm] {End};
	\draw [arrow] (start) -- (Jinit);
	\draw [arrow] (Jinit) -- (Jp);
	\draw [arrow] (Jp) -- (A0B0);
  \draw [arrow] (A0B0.south) -- +(0.0,-0.25) -| (sec1);
  \draw [arrow] (A0B0.south) -- +(0.0,-0.25) -| (sec2);
  \draw [arrow] (A0B0.south) -- +(0.0,-0.25) -| (secn);
  \draw [arrow] (sec1.south) -- +(0.0,-0.25) -| (damp);
  \draw [arrow] (sec2.south) -- +(0.0,-0.25) -| (damp);
  \draw [arrow] (secn.south) -- +(0.0,-0.25) -| (damp);
	\draw [arrow] (damp) -- (tol);
	\draw [arrow] (tol) -- node[anchor=east] {yes} (end);
	\draw [arrow] (tol) -- node[anchor=south] {no} +(-6.0,0.0) |- (Jp);

\end{tikzpicture}
\caption{The iterative parallel algorithm rapidly calculates the screening currents without loss of accuracy (simplified version for clarity).}
\label{f.method}
\end{figure}

The main improvement in the numerical method from \cite{pardo15SST} is an iterative process to reduce the computing time, while obtaining accurate results of the screening currents. We divide the coil cross-section in sectors containing several turns (figure \ref{f.sectors}), where the width of each sector in the radial direction is as close as possible to the tape width (except the last sector in each pancake, which contains the remaining tapes). The tapes in the sectors are divided into elements, where $J$ is assumed uniform. We start at the initial time $t_0$ with a known solution of the current density, $J_0$, for a given current $I_0$. After increasing the time by $\Delta t$, the routine calculates the change in current density $\Delta J$ due to the change in current $\Delta I$ (note that $t=t_0+\Delta t$, $J=J_0+\Delta J$ and $I=I_0+\Delta I$). The algorithm to obtain $\Delta J$ is outlined in figure \ref{f.method}. First, the algorithm sets uniform $\Delta J$ in the tape cross-section, $\Delta J=\Delta I/(wd)$. Then, the vector potential $\Delta A$ due to $\Delta J$ and the magnetic field $\bf B$ are calculated in all elements in each sector but taking only the contribution from $\Delta J$ outside the considered sector. We name these quantities as $\Delta A_e$ and ${\bf B}_e$.  Afterwards, the routine obtains $J$ at each sector ignoring the magnetic interaction with all the other sectors and taking $\Delta A_e$ and ${\bf B}_e$ as applied fields. The algorithm iterates until the maximum difference between iterations is below a certain tolerance. In order to ensure convergence of the iterative process, we apply a damping factor $0<k<1$ such that $\Delta J:=k\Delta J+(1-k)\Delta J_{\rm p}$, where $\Delta J_{\rm p}$ is $\Delta J$ at the previous iteration. The whole process is repeated for subsequent time increases $\Delta t$ until the desired final time is reached, obtaining the whole time evolution. At the beginning of the time evolution we take $t_0=0$, $I_0=0$ and we assume $J_0=0$, corresponding to the zero-field cool situation. The effect of the magnetic field generated by an external winding, such as a low-temperature superconducting outsert, can be simply taken into account by adding its contribution to $\Delta A_e$ and ${\bf B}_e$.

\begin{figure}[tbp]
\centering
{\includegraphics[trim=0 0 0 0,clip,width=8.5 cm]{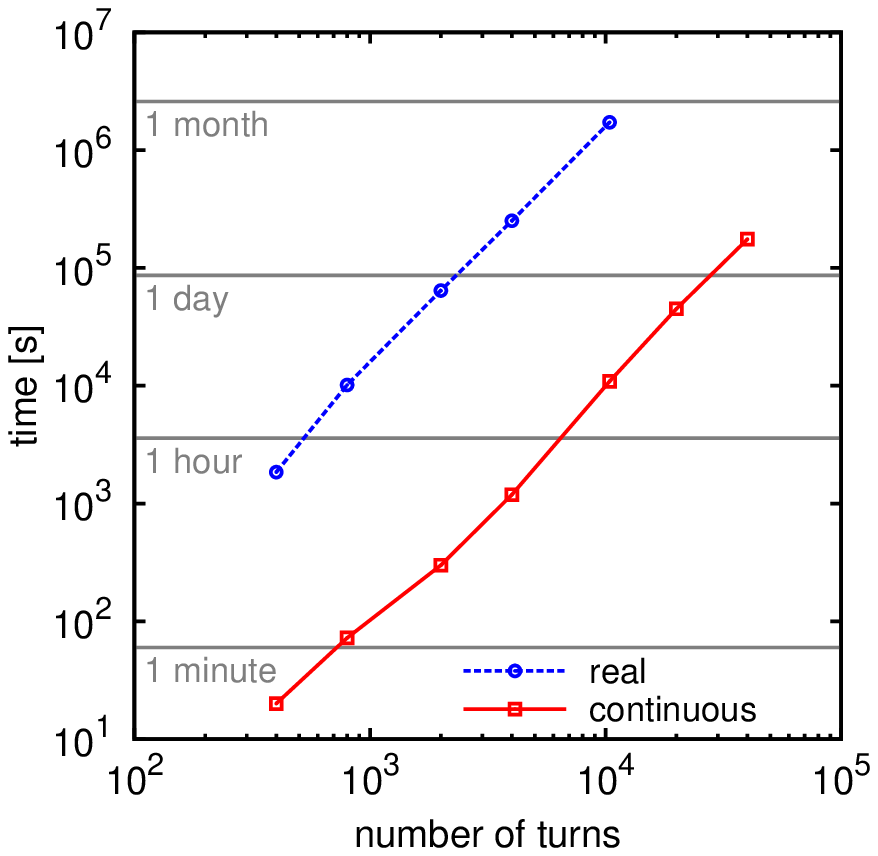}}
{\includegraphics[trim=0 0 0 0,clip,width=8.5 cm]{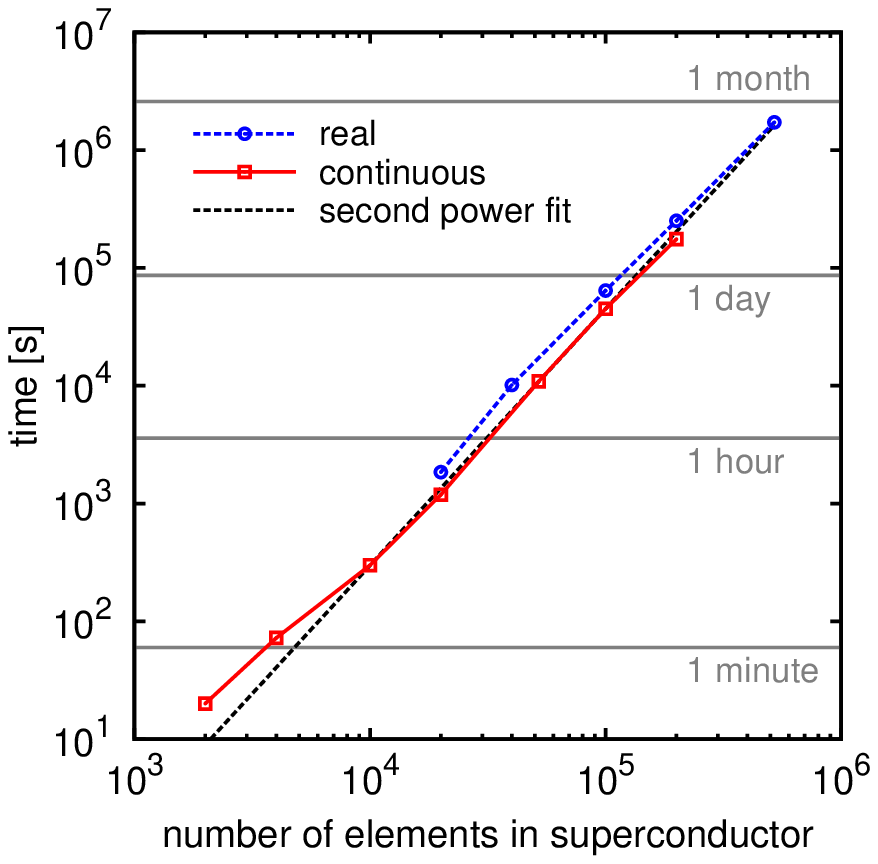}}
\caption{The self-programmed modeling tool computes coils up to 10400 turns for the real geometry and 40000 turns for the continuous approximation in feasible computing times, which scales as the second power of the number of turns or the number of elements.}
\label{f.time}
\end{figure}

For coils with many turns, this method greatly reduces the computing time. Since the computing time scales with the number of turns $N$ as $N^{\alpha}$ with $\alpha=2$ (figure \ref{f.time}); for a large number of turns, solving $m$ problems of $N/m$ turns takes only $1/m^{\alpha-1}$ of the original computing time. The number of iterations is kept low because the Screening Current Induced Field (SCIF) that the sector creates has a range of the order of the tape width, which corresponds to the sector size. In addition, this method can be easily parallelized, using up to as many tasks as sectors. The routine in this work uses more than 90 \% of the computing power of a 12-core computer for high number of turns. The sub-routine to calculate $\Delta A_e$ and ${\bf B}_e$ is also parallelized.

The computing time can be further reduced by using the continuous approximation for densely packed pancake coils \cite{prigozhin11SST,neighbour,zermeno13JAP}. This results in a reduction of the effective number of turns per pancake, in our case from 400 real turns to 40 effective turns.

Figure \ref{f.time} shows the computing time for both the real and continuous approximations. The computed case is the time evolution of the first ramp increase from 0 to the coil critical current. These calculations have been done in a computer with two 6-core Intel Xeon E5645 processors, where the required RAM is below 9 Gb for all cases. Computing times are similar to those in a standard 4-core (8 threads) processor Intel Core i7-4770K. Using 50 elements in the superconductor width and one in the thickness, the computing time for the real geometry with 400 and 1000 turns is around 30 min and 4.5 hours, respectively, and the computing time for 10000 turns is around 20 days (figure \ref{f.time}a). Using the continuous approximation greatly reduces the computing time (20 seconds, 103 seconds and 3 hours for 400, 1000 and 10000 turns, respectively), resulting in a factor around 100 reduction for high number of turns. This is consistent with the reduction of the effective turns by a factor 10 and the quadratic scaling of the computing time with the number of elements for high number of elements (figure \ref{f.time}b). The lower slope in figure \ref{f.time}b for the continuous approximation at low number of elements is due to a less efficient parallelization of the routine. With the continuous approximation, calculations for up to 40000 turns and 0.5 million elements in the superconductor have been effectuated in practical computing times. For all calculations, the computing time can be further reduced to half, if the mirror symmetry in the axial direction is assumed.

The DC critical current of a certain turn $u$ is calculated by taking a very slow ramp (10$^{-14}$ A/s). For this case, the general relation $E=-\partial_t A-(1/r)\partial_\varphi \phi_u$ becomes $E\approx -(1/r)\partial_\varphi \phi_u$, where $\phi_u$ is the scalar potential at turn $u$, $r$ and $\varphi$ are the radial and angular coordinates, respectively, and $\partial_\varphi \phi$ is uniform\footnote{For axial symmetry, ${\bf E}=E(r,z){\bf e}_\varphi$, where ${\bf e}_\varphi$ is the unit vector in the angular direction. Thus, ${\bf E}=-\partial_t{\bf A}-\nabla\phi$ implies $\partial_r\phi=\partial_z\phi=0$, and hence $\phi(r,\varphi,z)=\phi(\varphi)$. Since $E(r,z)$ does not depend on $\varphi$, $\partial_\varphi\phi$ is constant.}. The critical current of turn $u$ is that which follows $E=E_c$ in that turn. We take the critical current of the whole coil as that of the weakest turn. Comparing to other methods \cite{pitel13SST,gomory13IES,zermeno15SST}, the method in this article takes the precise magnetization currents into account. In addition, it allows to calculate the coil DC current-voltage curve with non-uniform power-law exponent, $n$, and $n(B,\theta)$.

\section{Results and discussion}

This section presents the results for the screening currents, Screening Current Induced Field (SCIF) and instantaneous AC loss for the coil with parameters in table \ref{t.geom}. The SCIF at the remnant state and the critical current is also presented for coils from 2 to 100 pancakes. The number of elements per turn (real or effective) is between 50 and 200, being the higher number of elements for the loss computations. The ramp rate is for 2 A/s for figures \ref{f.Jireal},\ref{f.Jidet},\ref{f.Jicont},\ref{f.Jrcont} and 0.2 A/s for the rest, being the qualitative behavior independent on this parameter.


\subsection{Current density}

\begin{figure}[tbp]
\centering
{\includegraphics[trim=125 3 153 8,clip,width=8.5 cm]{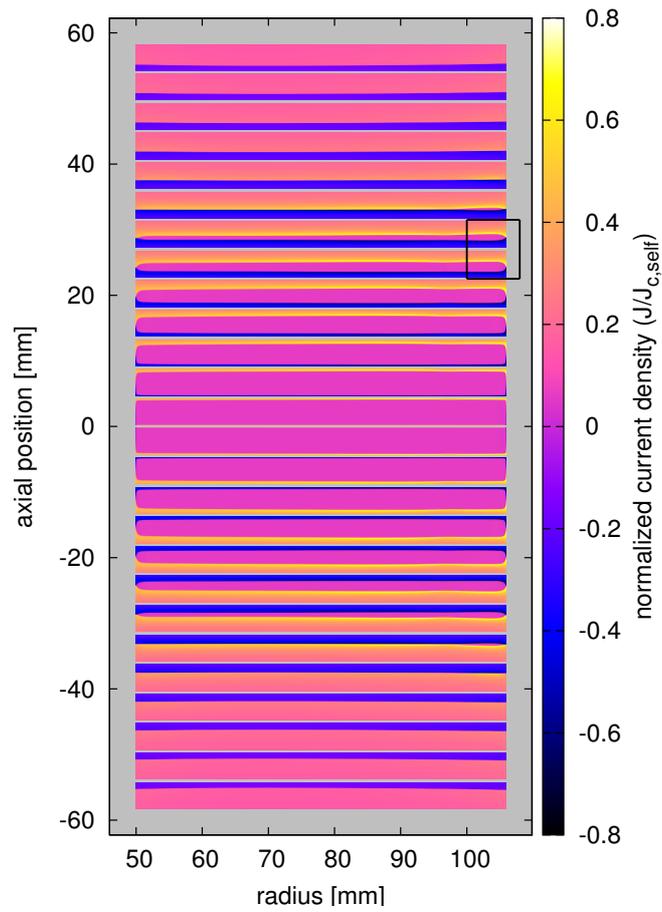}}
\caption{The coil presents important screening currents at the end of the (first) increasing ramp. Results are for the real geometry; the pancakes cross-section is depicted as continuous for clarity. A detail of the region within the black frame is shown in figure \ref{f.Jidet}.}
\label{f.Jireal}
\end{figure}

\begin{figure}[tbp]
\centering
{\includegraphics[trim=55 2 72 10,clip,width=8.5 cm]{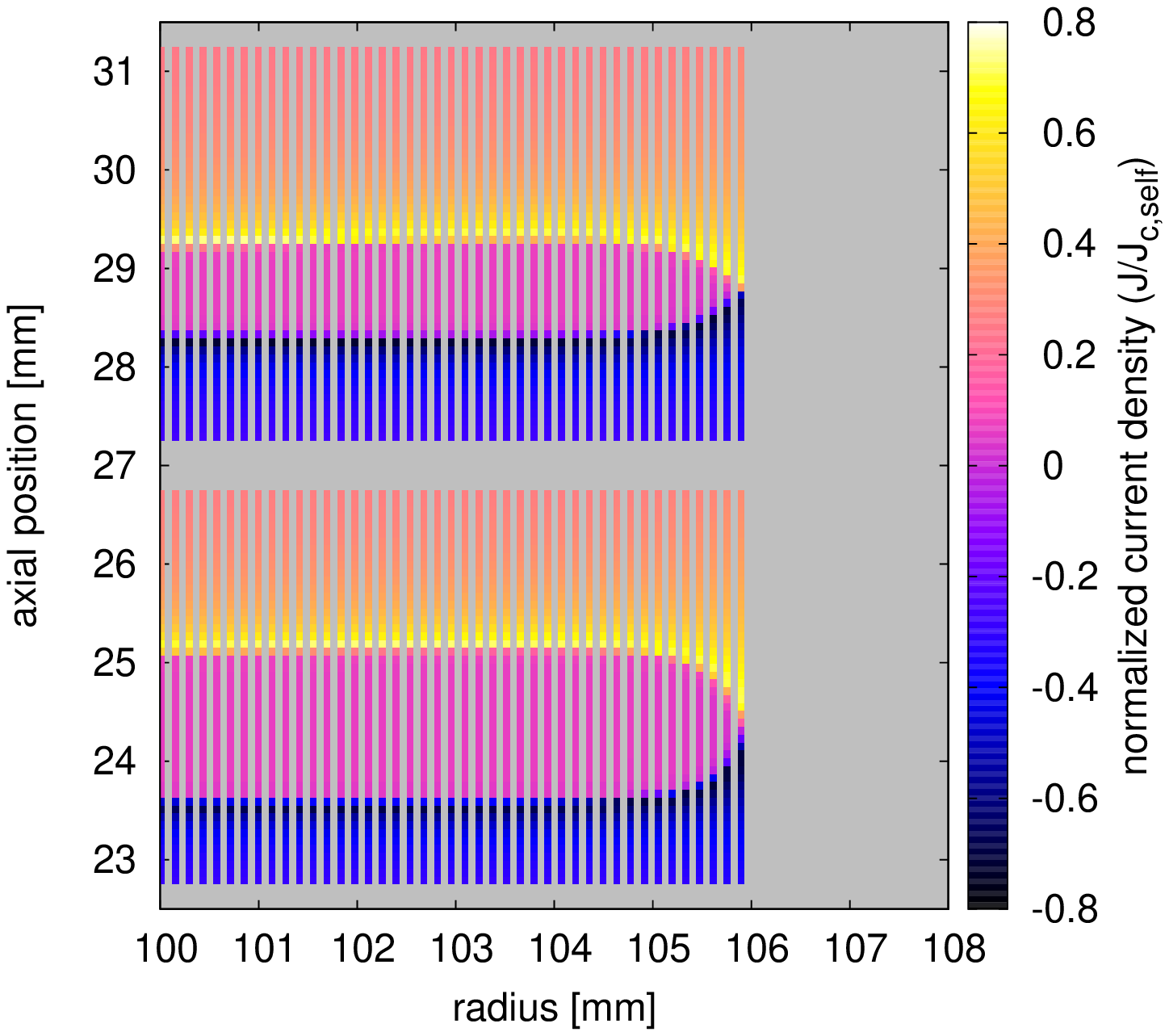}}
\caption{Detail of the screening currents in figure \ref{f.Jireal} (area within the black frame there). The superconducting-layer thickness is expanded in the plot for visibility (computations use realistic 2 $\mu$m thickness).}
\label{f.Jidet}
\end{figure}

Figure \ref{f.Jireal} shows the current density at the peak of the charge cycle in the coil of 400$\times$26 turns for the detailed model (coil cross-section in figure \ref{f.sketch} and current signal in figure \ref{f.signal}). For this case, we assumed a power-law exponent of 30. The computed situation is for a peak current of 226 A, which represents 76 \% of the critical current. A zoom of the area marked by a black frame is in figure \ref{f.Jidet}, showing the individual turns. The largest screening currents are at around 3/4 of the coil height (6th and 7th pancakes from the top). This is because at the end turns the large magnetic field, around 8 T, substantially suppresses $J_c$; at the central turns, the radial field is much lower than the penetration field of the pancake, and hence screening currents are small. The screening currents are roughly the same in all turns of each pancake, except at the 8-10 turns the closest to the inner and outer radius (see figure \ref{f.Jidet}). At the 3 pancakes the closest to the end, variations of the screening currents in the radial direction are due to variations in the magnetic field, specially its radial component.

\begin{figure}[tbp]
\centering
{\includegraphics[trim=125 3 153 8,clip,width=8.5 cm]{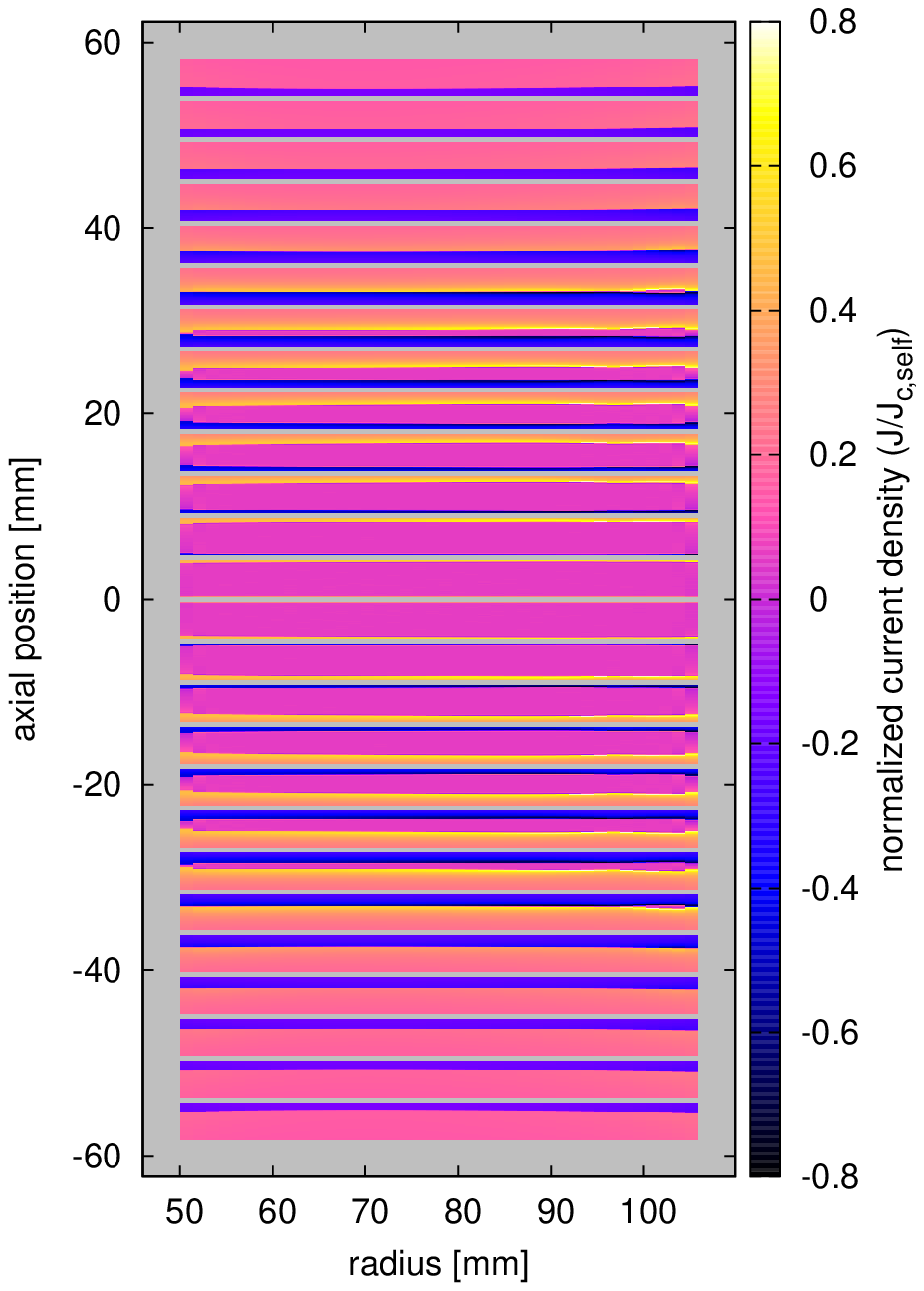}}
\caption{The screening currents for the continuous approximation agree with those for the real geometry (same situation as figure \ref{f.Jireal}).}
\label{f.Jicont}
\end{figure}

The continuous approximation reproduces the same current density as the real geometry (see figures \ref{f.Jireal} and \ref{f.Jicont}), except of a certain averaging in the radial direction for the 8-10 turns the closest to the inner or outer radius. The results could be improved by using thinner effective turns close to the inner and outer radius.

\begin{figure}[tbp]
\centering
{\includegraphics[trim=2 106 8 95,clip,width=15 cm]{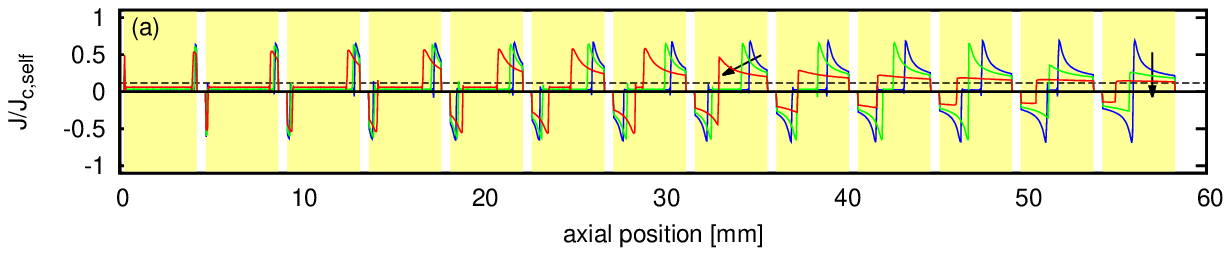}}
{\includegraphics[trim=2 106 8 95,clip,width=15 cm]{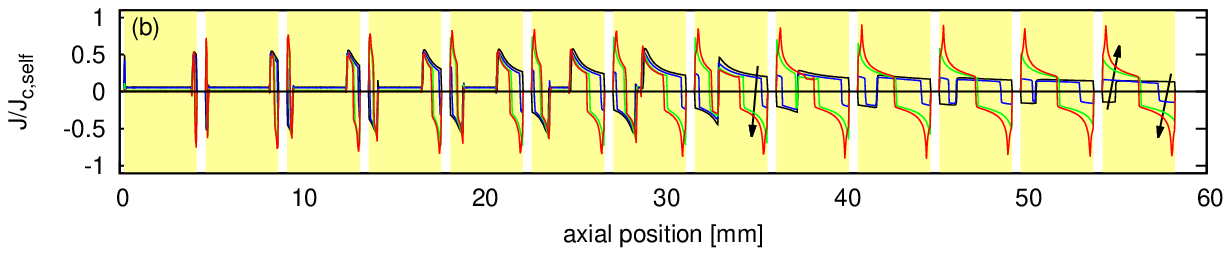}}
{\includegraphics[trim=2  88 8 95,clip,width=15 cm]{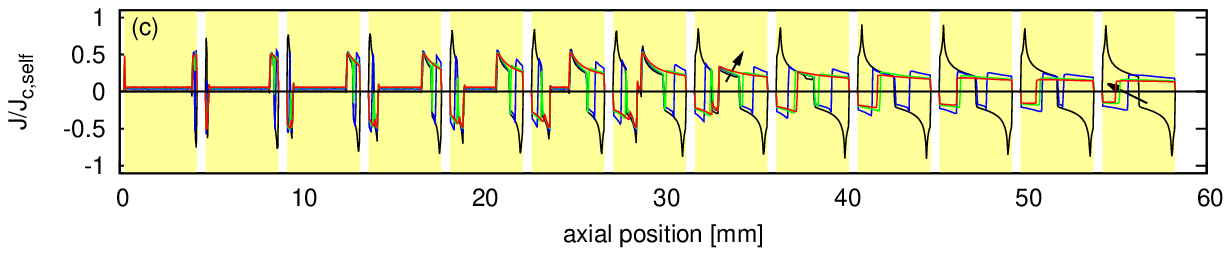}}
\caption{The current density in the turns at the mid radius show the main features of the current penetration process (continuous approximation). The situation for each curve corresponds to those in figure \ref{f.P}a with labels 11,12,13 (a), 13,14,15,16 (b), and 16,17,18,19 (c) in the arrows direction. The horizontal dash line in (a) represents the current density at the end of the ramp if magnetization currents are not present.}
\label{f.Jcen}
\end{figure}

\begin{figure}[tbp]
\centering
{\includegraphics[trim=2 88 8 95,clip,width=15 cm]{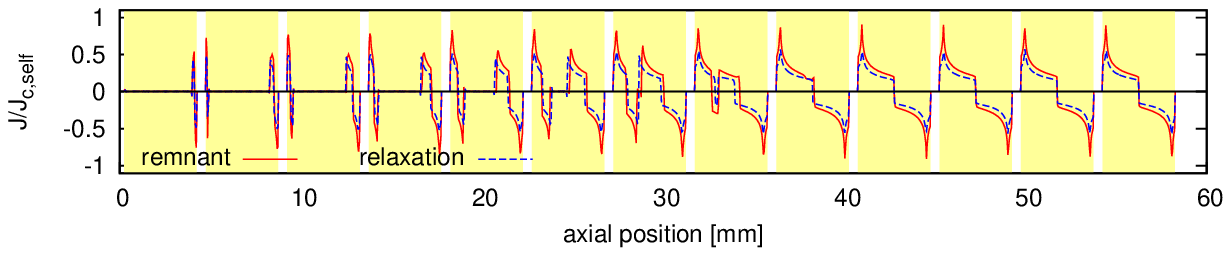}}
\caption{Relaxation after the remnant state (situation of figure \ref{f.signal}b) reduces the screening currents (results for the turns at the mid radius for the continuous approximation).}
\label{f.Jcenrel}
\end{figure}

The magnitude and current penetration process can be better appreciated in the line plots of the current density at the mid radius, 77.95 mm (see figures \ref{f.Jcen} and \ref{f.Jcenrel} for the continuous approximation).

The typical penetration process of the initial ramp can be seen from pancake 3 from the coil end in figure \ref{f.Jcen}a. First, the current penetrates with $|J|\approx J_c$ (critical region) from both sides with a sub-critical region in between (profile 11). The magnetic field dependence of $J_c$ causes a variation of $J_c$ in the axial direction. The radial field vanishes at the sub-critical region (region with $|J|$ substantially smaller than $J_c$), and hence $|J|$ is the largest at the front of critical current density. The current does not vanish in the sub-critical region. By further increasing the current $I$, the critical region expands until it fully penetrates the tape (profile 12). Afterwards, $J_c$ decreases due to the increase of $B_r$ and the zone with negative current is reduced, caused by the increase of $I$ (profile 13).

For the decreasing ramp, we also analyze pancake 3 from the coil end in figure \ref{f.Jcen}b. A zone with $|J|\approx J_c$ is created (reverse critical region) with $J$ of the opposite sign from the preexisting (profile 14). The penetration process continues until the sample saturates, while the current density increases in the reverse critical region (profile 15); this is caused by the increase of $J_c$ due to the decrease of the radial magnetic field. After full penetration, $J$ further increases (profile 16). An example of partial penetration of the reverse critical region is pancake 6 from the end, where there remains a zone with $J$ induced at the initial ramp enclosed by the reverse critical region.

\begin{figure}[tbp]
\centering
{\includegraphics[trim=125 3 153 8,clip,width=8.5 cm]{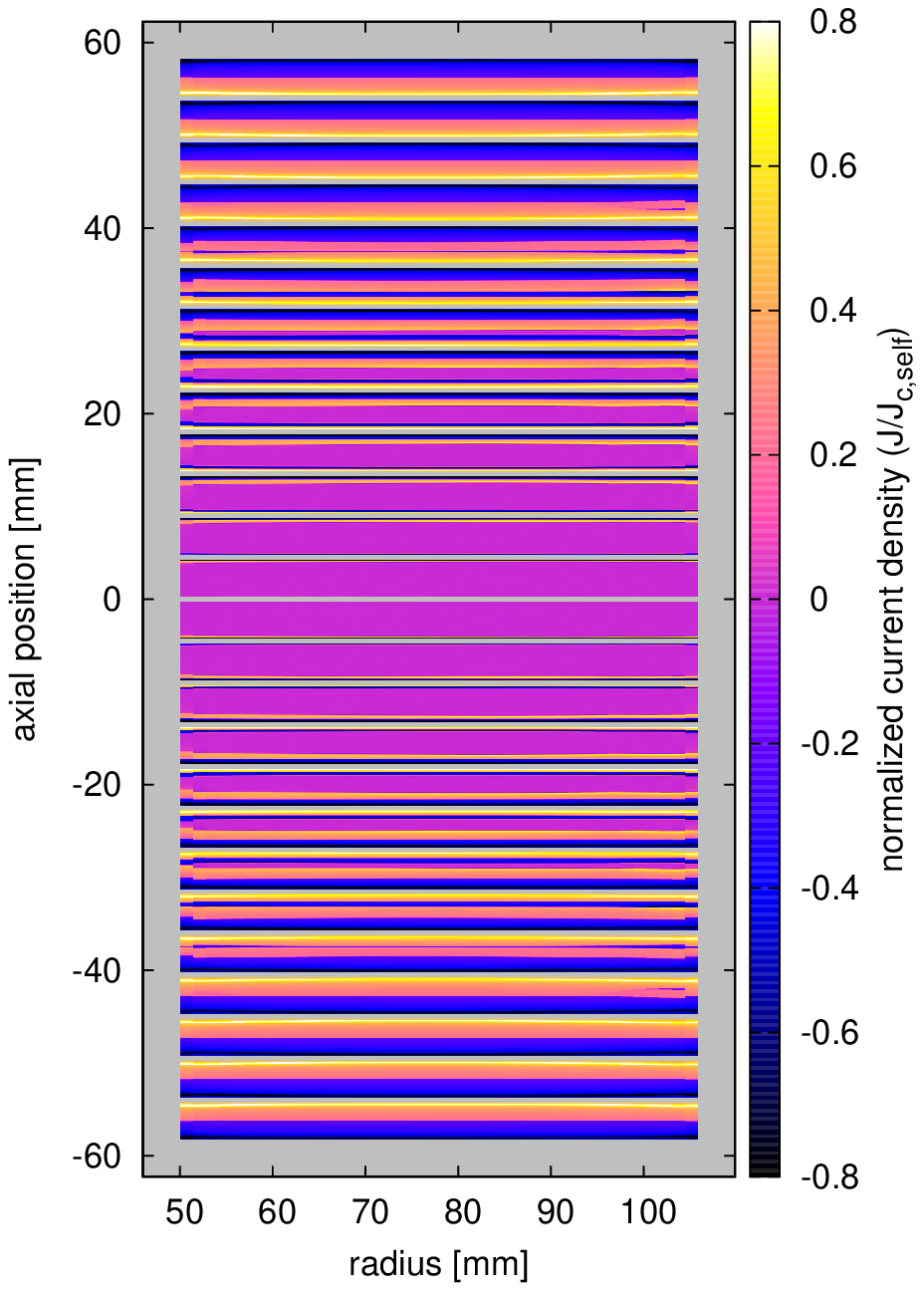}}
\caption{The screening currents are the most important at the remnant state (results for the continuous approximation).}
\label{f.Jrcont}
\end{figure}

At the remnant situation, or end of the discharge ramp, screening currents are more important than at the maximum current (figures \ref{f.Jrcont} and \ref{f.Jcen}b). The reasons are the following: first, the transport current in the charge curve partially depletes the region of negative current; second, the radial field for the remnant case is reduced down to the self-field of the pancake ($\approx$ 3 T at the center of the top pancake, compared to $\approx$ 8 T), resulting in higher $J_c$. The maximum screening currents are at the top 4 pancakes, since there the variation of radial field from the peak current is the largest.

If the current is switched off at the remnant state, the average current density decreases due to relaxation caused by the finite power-law exponent. For our case, relaxation after 10$^7$ seconds appreciably reduces the screening currents, although they are still important (figure \ref{f.Jcenrel}).

\subsection{Screening Current Induced Field}

In the following, we analyze the SCIF at the bore center. We consider the 10400 turns coil of table \ref{t.geom} for both cyclic ramp-up and ramp-down excitations (section \ref{s.cyc}) and relaxation of the remnant state (section \ref{s.rel}). Section \ref{s.remIc} describes the effect of the number of pancakes, or coil size, on the SCIF at the remnant state.

\subsubsection{Cyclic charge and discharge}
\label{s.cyc}

\begin{figure}[tbp]
\centering
{\includegraphics[trim=2 2 10 10,clip,width=11 cm]{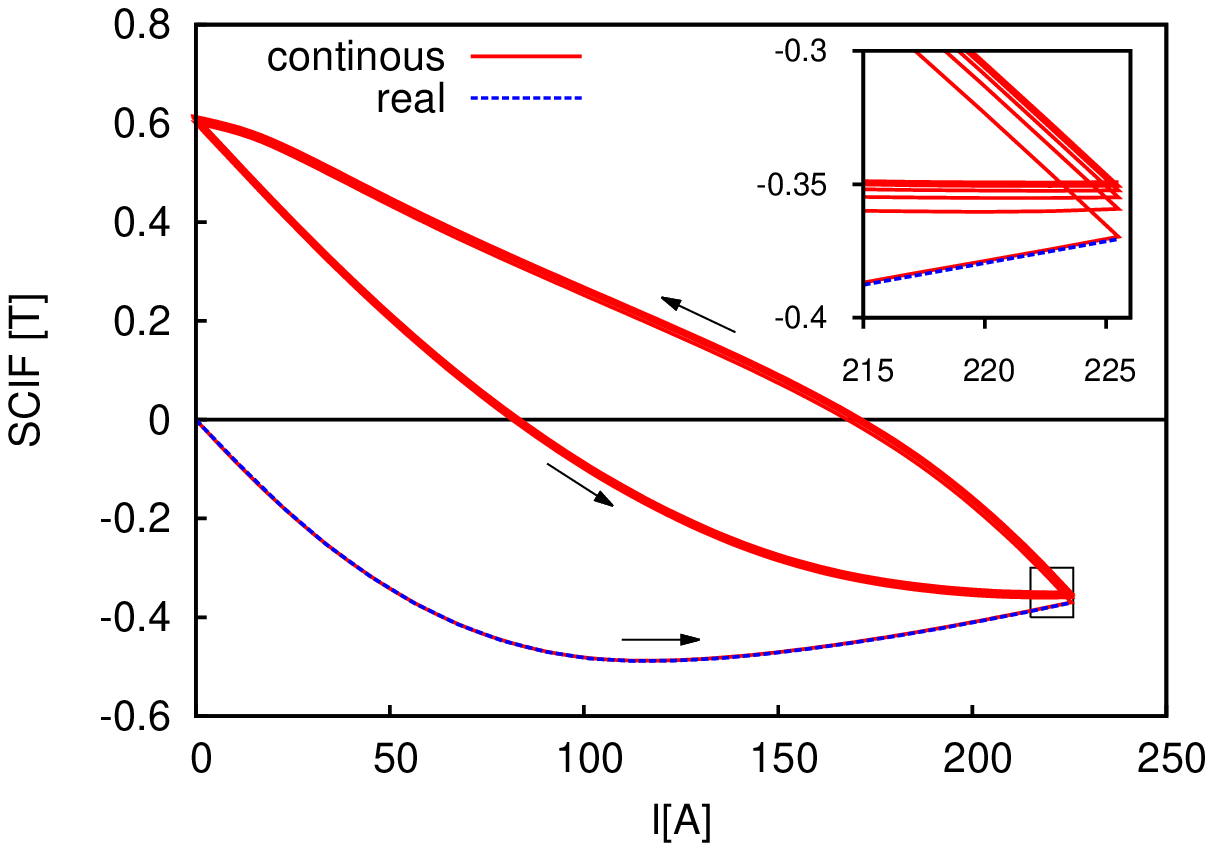}}
\caption{The screening current induced field (SCIF) taking the real geometry and continuous approximation agree with each other. The inset shows a detail of the area within the black frame.}
\label{f.SCIF}
\end{figure}

For the initial ramp, taking the continuous approximation into account reproduces the curve for the real situation with a maximum deviation of 0.2 \% of the SCIF at the peak, as shown in figure \ref{f.SCIF}. Then, the continuous approximation can be used to predict the SCIF and optimize the magnet design regarding this aspect.

For the cyclic charge and discharge (figure \ref{f.signal}b), the SCIF at the first charge is different than that at the following charges (see figure \ref{f.SCIF}) due to the residual current density from the previous discharge. The magnitude of the SCIF slowly decreases with the number of charges until it reaches a steady-state cycle (see inset of figure \ref{f.SCIF}). This is caused by the finite power-law exponent of the $E(J)$ relation, which allows relaxation of the residual screening currents created at the first charge. At the initial charge, the decrease of the SCIF with increasing the current at high currents is due to both a depletion of negative current caused by transport current and the suppression of $J_c$ as a consequence of the increase of the magnetic field.

The SCIF in the coil of figure \ref{f.sketch} is relatively large compared to the maximum generated field (the SCIF are -0.37 T and 0.61 T at the peak and remnant state, respectively, compared to 14.9 T generated magnetic field at the peak). This may be an issue for applications that require generated magnetic fields of high quality (homogeneity and stability), such as MRI or NMR.

\subsubsection{Relaxation in remnant state}
\label{s.rel}

\begin{figure}[tbp]
\centering
{\includegraphics[trim=2 2 10 10,clip,width=11 cm]{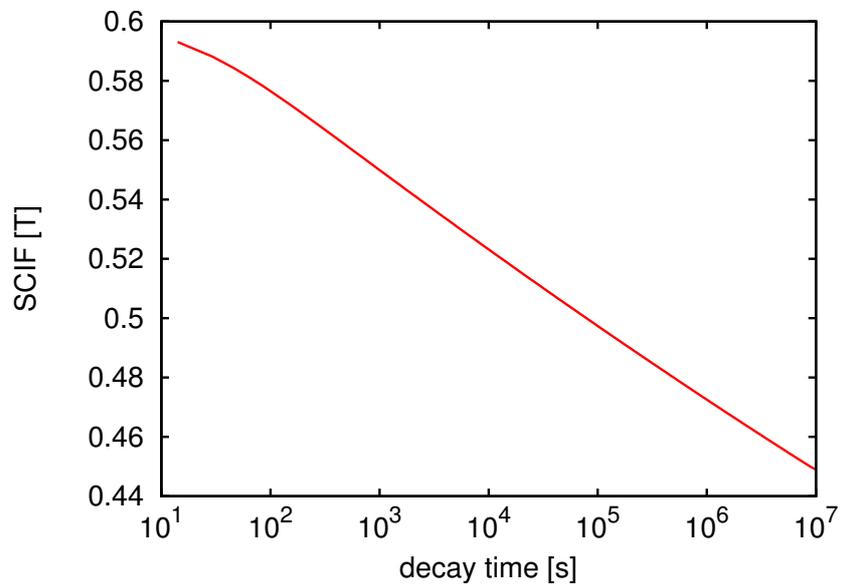}}
\caption{The Screening Current Induced Field (SCIF) decays in time roughly exponentially.}
\label{f.Bcdecay}
\end{figure}

After one single charge and discharge (case in figure \ref{f.signal}a), the SCIF decreases with time due to the smooth $E(J)$ relation (see figure \ref{f.Bcdecay}). At high relaxation times, the SCIF decreases exponentially, which is a typical behavior for power-law $E(J)$ relations \cite{brandt95RPP}. The change in the SCIF is 19, 65, 101, 122, 138 mT in 1 minute, 1 hour, 1 day, 1 week and 1 month, respectively, representing 0.13, 0.44, 0.68, 0.82, 0.93 \% of the generated magnetic field at the peak of the current. 

\subsubsection{Dependence on the number of pancakes}
\label{s.remIc}

To study the dependence of the SCIF on the coil size, we calculated the SCIF at the remnant state for coils from 2 to 100 pancakes, where each pancake contains 400 turns (40000 turns in total for the largest coil). For all coils, we considered the case that the peak current corresponds to the coil critical current.

\begin{figure}[tbp]
\centering
{\includegraphics[trim=4 2 20 7,clip,width=10 cm]{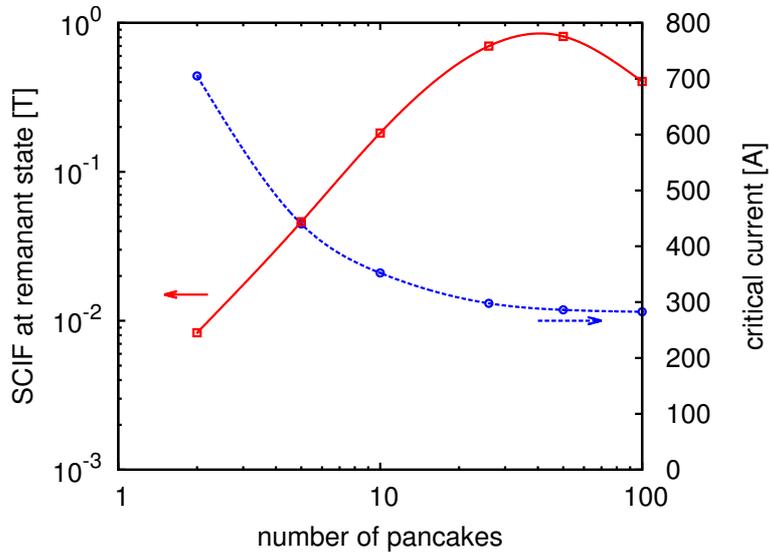}}
\caption{The Screening Current Induced Field (SCIF) at the remnant state presents a peak with increasing the number of pancakes, while the critical current saturates.}
\label{f.rempan}
\end{figure}

The critical current decreases with the number of pancakes until it saturates for a high number of pancakes (see figure \ref{f.rempan}; data in table \ref{t.Ic}). For all cases, the turn limiting the critical current is located at the top pancake. This behaviour is caused by the fact that $J_c$ is very sensitive to the radial field and this is the highest at the top pancakes. With increasing the number of pancakes, the radial field increases until it saturates for long coils. The position of the maximum radial field at the top pancake moves to inner radius with increasing the number of pancakes. As a consequence, the turn with the lowest critical current at the top pancake moves accordingly (table \ref{t.Ic}).

The SCIF at the coil center increases with the number of pancakes until around 40 pancakes and then it decreases for higher number of pancakes (see figure \ref{f.rempan}). The reason is that for high number of pancakes, the height-over-width aspect ratio of the winding is large. For this geometry, the radial magnetic field is low except at the coil ends. As a consequence, the screening currents are only important close to the coil ends. Then, with increasing the coil height, the distance between the pancakes contributing to the SCIF and the coil center increases, and hence the SCIF decreases.

\begin{table}
\begin{center}
\begin{tabular}{llll}
\hline
\hline
{\bf number of} & {\bf number of } & {\bf  turn with lowest } & {\bf  critical current}\\
{\bf pancakes}	& {\bf  turns			} & {\bf  critical current	} & {\bf  [A]}\\
\hline
2 & 800 & 310 & 705\\
5 & 2000 & 150 & 440\\
10 & 4000 & 140 & 352\\
26 & 10400 & 130 & 298\\
50 & 20000 & 130 & 286\\
100 & 40000 & 130 & 283\\
\hline
\hline
\end{tabular}
\caption{Critical current of the weakest turn for coils with several number of pancakes (the rest of the parameters are the same as those in table \ref{t.geom}). The turn with the lowest critical current is always at the end pancakes.}
\label{t.Ic}
\end{center}
\end{table}


\subsection{AC loss}
\label{s.loss}

\begin{figure}[tbp]
\centering
{\includegraphics[trim=3 18 17 5,clip,width=10 cm]{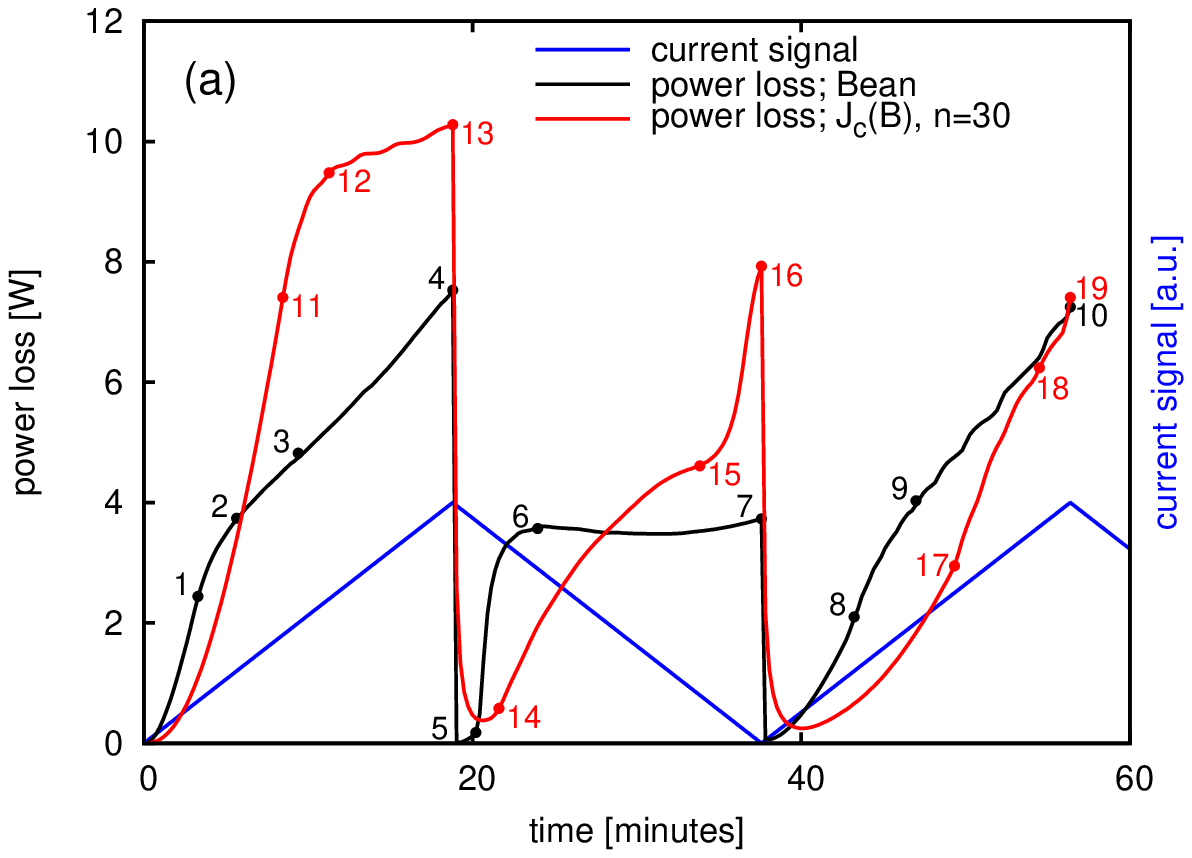}}
{\includegraphics[trim=3  2 17 5,clip,width=10 cm]{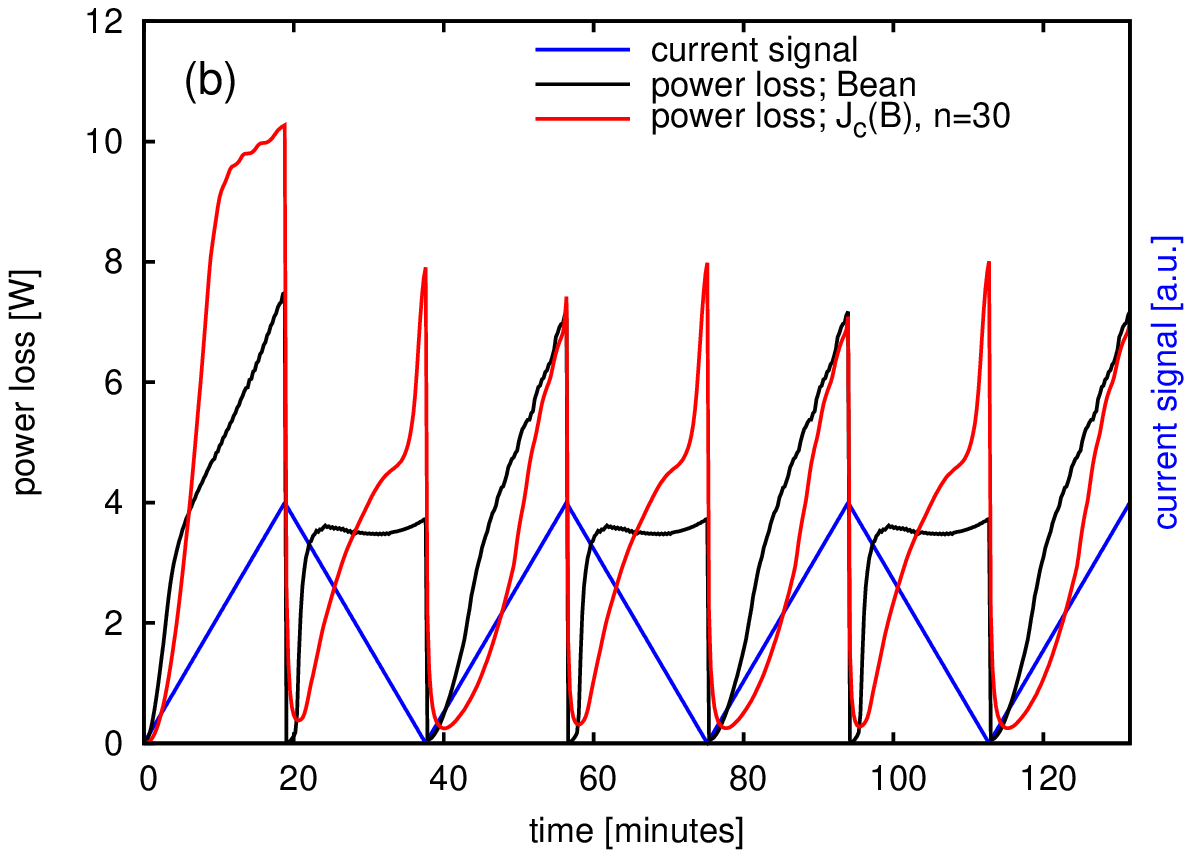}}
\caption{The instantaneous AC loss for the taken $J_c(B,\theta)$ dependence and power-law exponent $n=30$ is qualitatively very different than that for constant $J_c$ and $n=1000$ (critical-state model or ``Bean" case). The numbers in (a) represent the key events in the current penetration process (see figures \ref{f.Jcen},\ref{f.Jcenbean} and text). The peak current is $I_m$=226 A.}
\label{f.P}
\end{figure}

In order to understand the causes of the time dependence of the power AC loss, we calculated both the same situation as figure \ref{f.SCIF} (anisotropic magnetic-field dependence of $J_c$ from equations (\ref{IcBth}) and (\ref{wB}) and power-law exponent 30) and that for constant $J_c$ and power-law exponent 1000, the latter corresponding to the Bean's critical state model. For the Bean's situation, we took $J_c=I_c/(wd)$, where $I_c$ is the coil critical current for the magnetic-field dependent $J_c$. Although the loss per cycle is of the same order of magnitude (7.72 and 6.81 kJ for the Bean's and $J_c(B,\theta)$ cases, repectively), the instantaneous loss differs substantially and presents different qualitative behavior for the ramp decreases (figure \ref{f.P}).

\subsubsection{Bean's case}

\begin{figure}[tbp]
\centering
{\includegraphics[trim=2 106 8 95,clip,width=15 cm]{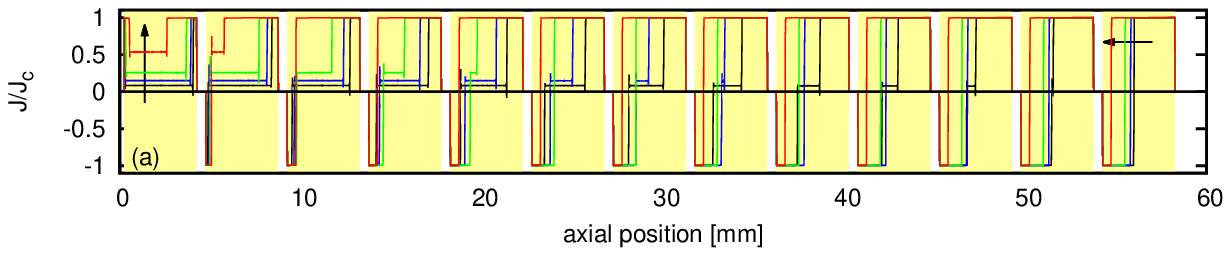}}
{\includegraphics[trim=2 106 8 95,clip,width=15 cm]{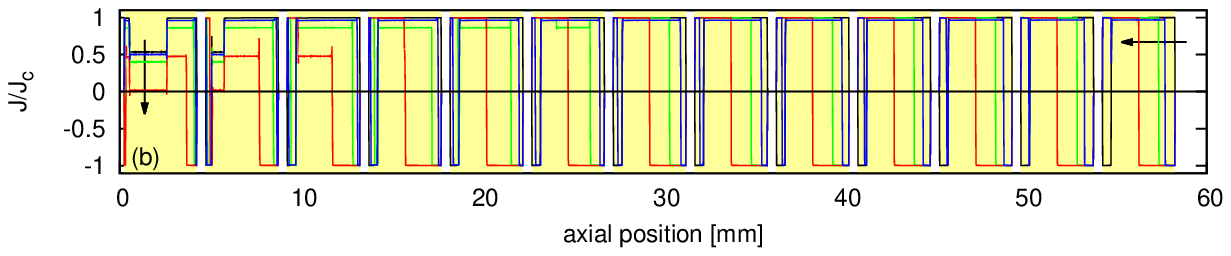}}
{\includegraphics[trim=2  88 8 95,clip,width=15 cm]{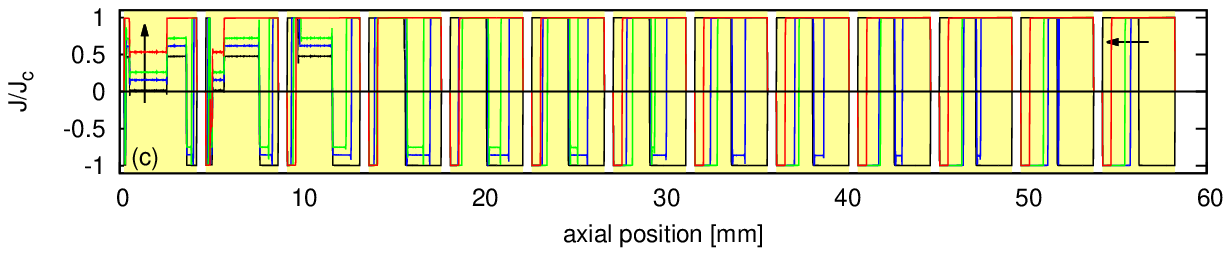}}
\caption{Current density in the turns with mid radius for constant $J_c$ and power-law exponent 1000, corresponding to the critical-state model. The situation for each curve is the same as that of the points in figure \ref{f.P}a with labels 1,2,3,4 (a), 4,5,6,7 (b), 7,8,9,10 (c), following the arrow direction.}
\label{f.Jcenbean}
\end{figure}

For the Bean's case, the loss at the initial ramp increases as $I^2$ until it reaches an inflexion point (point 1 in figure \ref{f.P}). The $I^2$ dependence is explained by the partial penetration by $J_c$, as has been shown for slabs in applied magnetic field \cite{schwerg12IES}. The inflexion point corresponds to the penetration of the the two pancakes from the top (figure \ref{f.Jcenbean}a profile 1). Afterwards, the loss increases slower with roughly a linear dependence. This linear dependence roughly starts when the top six pancakes are fully penetrated (figure \ref{f.Jcenbean}a profile 2), which generate most of the AC loss. Again, the $I$ dependence is consistent with the behavior of a fully penetrated slab under applied magnetic field. For large enough current, the loss rises faster than a liner dependence due to the direct effect of the transport current, which creates dissipation due to dynamic magneto-resistance.

At the beginning of the ramp decrease, the loss drops, since the time derivative of the current vanishes (figure \ref{f.P}). Later, when the critical region with reverse current density meets the sub-critical region (figure \ref{f.Jcenbean}b profile 5), the loss experience a sharp increase (point 5 in figure \ref{f.P}). This is because a small change in $I$ causes penetration of the sub-critical region, becoming exposed to the magnetic flux ``external" to the pancake, and hence that region also experiences AC loss. This penetration of the sub-critical region is slower for pancakes further from the ends, since the increase of the ``external" magnetic field is smaller for the same current increase. Therefore, when a sufficient amount of pancakes are penetrated, the loss starts to decrease (figure \ref{f.P}). At a certain time, the loss increases again because by decreasing $I$, the magnetization currents at the fully penetrated pancakes increase; since by decreasing $I$, the current constrain allows a higher volume for the magnetization currents (compare profiles 6 and 7 in figure \ref{f.Jcenbean}b for the 7 upper pancakes).

At the second ramp increase, the qualitative behavior is the same as for the initial curve (figure \ref{f.P}). The only difference is that the quadratic dependence, appearing during partial penetration of the newly created critical region, expands to higher $I$ (up to point 8 in figure \ref{f.P}). That is because for this case the local change in $J$ during critical zone penetration is $2J_c$, instead of only $J_c$ in the initial curve (figures \ref{f.Jcenbean}a and \ref{f.Jcenbean}b).

\subsubsection{Anisotropic magnetic field dependence of $J_c$}

For the $J_c(B,\theta)$ dependence, the initial curve is qualitatively the same as for the Bean's case (figure \ref{f.P}). However, the interval with quadratic increase is larger (interval up to point 11). The quasi-linear increase that follows presents a lower slope because $J_c$ decreases with increasing $I$ due to the increase of the radial magnetic field. The loss at the current peak is higher for $J_c(B,\theta)$ because the local $J_c$ is higher in the pancakes 2 to 7 from the end, producing higher magnetization loss. 

The loss in the decreasing ramp looks very different from the Bean's case (figure \ref{f.P}). Although it also starts with an increase after the penetration of the first pancake (point 14 in figure \ref{f.P} and profile 14 in figure \ref{f.Jcen}b), this increase presents a lower rate because when decreasing $I$, the decrease of $B_r$ causes an increase of $J_c$ of the reverse critical region. As a consequence, penetrating sub-sequent pancakes requires a larger variation of $I$ (or magnetic field), due to the larger shielding capability of the reverse currents. Between points 15 and 16 in figure \ref{f.P}, there is a sharp increase of the loss. In this interval, there is practically no further penetration by reverse current but the current density at the fully penetrated pancakes sharply increases due to the decrease of the local $B$ and the sharp $J_c(B)$ dependence at low $B$ (current profiles in \ref{f.Jcen}b).

The loss in the second ramp increase is qualitatively similar to the first ramp, except that the region with quadratic increase expands to a much larger interval for the same reasons as for the Bean's case. The loss at the end of the second ramp increase is lower than at the end of the first. The cause is that at the end of the second ramp, the newly induced critical region, where $E$ is significant, is smaller than that in the initial curve [except for the five pancakes from the top, where $J$ is the same (figures \ref{f.Jcen}a and \ref{f.Jcen}c)]. In the other pancakes, the small deep in $|J|$ in the negative $J$ region shows the border between both critical regions (profile 19 of figure \ref{f.Jcen}c). For ramp increases after the first, the loss at the peak slightly decreases with the number of cycles. The reason is the decay of the frozen current density at the first ramp, caused by the finite power-law exponent. This also causes a time delay of the loss minimum at the end of all ramps. 


\section{Conclusions}

This article presented detailed modelling of the screening currents and the screening current induced field (SCIF) of coils with more than 10000 turns for the real geometry and 40000 turns for the continuous approximation. We also analyzed the instantaneous power loss and the causes that create it. Since we used an anisotropic field dependence of $J_c$ extracted from measurements at 4.2 K from self-field up to 30 T, the results are representative for high-field magnets. The qualitative behavior is valid for any winding containing densely packed pancake coils, such as SMES.

The calculations used the MEMEP numerical method \cite{pardo15SST}. In order to achieve feasible computing times while keeping the same accuracy, we developed a fast iterative routine that we efficiently parallelized. The same concept might be applied to conventional finite-element methods (FEM).

The results show that the continuous approximation is suitable to calculate the screening currents and the SCIF, while further reducing the computing time. For a cyclic charge and discharge process, the stationary loop is achieved after several cycles due to the assumed power-law $E(J)$ relation, which causes relaxation of the magnetization currents induced at the first ramp. With increasing the number of pancake coils, the SCIF at the remnant state presents a peak at around 40 pancakes, while the coil critical current monotonically decreases until it saturates. Comparing to the results with a realistic $J_c(B,\theta)$, taking $J_c$ from the coil critical current results in a loss per cycle of the same order of magnitude (only 13 \% overestimation) but the instantaneous loss is very different. Saturation of the magnetization currents at the end pancakes, where the AC loss is the largest, causes that the maximum AC loss at the first ramp increases with increasing the critical current density.

This article have shown that the presented model can accurately calculate the SCIF in practical computing times for coils with any number of turns used in real windings. Such modeling could also predict the power loss in the charge and discharge process and the relaxation of the SCIF after one charge or discharge cycle. Experimental data on the anisotropic magnetic field dependence of the power-law exponent may be used to refine the calculations.


\section*{Acknowledgements}

The research leading to these results has received funding from the European Union Seventh Framework Programme [FP7/2007-2013] (grant NMP-LA-2012-280432) and the Structural Funds of EU (grant ITMS 26240220088)(0.5). The authors acknowledge the use of resources provided by the SIVVP project (ERDF, ITMS 26230120002).




\bibliographystyle{unsrt}

\end{document}